\documentstyle[preprint,aps]{revtex}
\begin{document}
\draft
\title {\bf EFFECT OF MAGNETIC FIELD ON THE STRANGE STAR}
\author{S. Chakrabarty$^1$ and P.K. Sahu$^2$}
\address{$^{1}$Department of Physics, University of Kalyani, Kalyani 741 235,
India;\\
E-mail: somenath@klyuniv.ernet.in.}
\address{$^{2}$Theory Group, Physical Research Laboratory, Ahmedabad 380 009,
India;\\
E-mail: pradip@prl.ernet.in.}
\maketitle
\begin{abstract}
We study the effect of a magnetic field on the strage quark matter and apply
to strange star. We found that the strange star becomes more compact in
presence of strong magnetic field.
\vskip 0.4in
\end{abstract}
There are several different scenarios for estimation of the magnetic
field strength at the surface of neutron star. These are: theoretical
models of pulsar emission\cite{rud} the accretion flow in the binary X-ray
sources\cite{gho} and observation of features in the spectra of pulsating X-ray
sources which have been interpreted as cyclotron lines\cite{tru}. In a sample
of more than 300 pulsars the range of values of the surface magnetic field
strength runs into the interval 10.36 $\le$ log($B$) (G) $\le$ 13.33\cite{man}.
\par
Very recently, several authors\cite{dun} have proposed two different physical
mechanism leading to an amplification of some initial magnetic field in
a collapsing star. Fields as strong as $B \sim 10^{14}~~\sim 10^{16} $G
, or even more, might be generated in new born neutron stars. In the
interior
of neutron star, it probably reaches $\sim 10^{18}$G. Therefore, it is
advisable to study the effect of strong magnetic field on compact neutron
stars.
\par
There are strong reason for believing that the hadrons are composed of quarks
, and the idea of quark stars has already been existed for about twenty years.
If the neutron matter density at the core of neutron stars exceeds a few times
normal nuclear density ($\> 3n_{0}, n_{0}=0.15~fm^{-3}$) a deconfining phase
transition to quark matter may take place. As a consequence, a normal neutron
star will be converted to a hybrid star with an infinite cluster of quark
matter core and a crust of neutron matter. In 1984, Witten suggested that
strange matter, e.g., quark matter with strangeness per baryon of order unity,
may be the true ground state\cite{wit}. The properties of strange matter at
zero
pressure and zero temperature were subsequently examined, and it was found
that strange matter can indeed be stable for a wide range of parameters in
the strong interaction calculations\cite{far}. Therefore, at the core,
the strange
quarks will be produced through the weak decays of light quarks (u and d
quarks) and ultimately a chemical equilibrium will be established among the
participants. Since, the strange matter is energitically favourable over
neutron
matter, there is a possibility that whole star may be converted to a strange
star.
\par
For a constant magnetic field along the z-axis ($\vec A =0, \vec H=H(z)=H
=constant$), the single energy eigenvalue is given by\cite{lan}
\begin{equation}
\varepsilon_{p,n,s}=\sqrt{{p_i}^2+{m_i}^2+q_iH(2n+s+1)}
\end{equation}
where n=0, 1, 2, ..., being the principal quantum numbers for allowed Landau
levels, $s=\pm1$ refers to spin up(+) and down(-) and $p_i$ is the component of
particle(species i) momentum along the field direction. Setting $2n+s+1=2\nu$,
where
$\nu = 0, 1, 2...,$ we can rewrite the single particle energy eigenvalue
in the following form
\begin{equation}
{\varepsilon_{i}}=\sqrt{{p_i}^2+{m_i}^{2}+2\nu q_iH}
\end{equation}
Now, it is very easy to see that $\nu=0$ state is singly degenerate,
whereas, all other states with $\nu\not=0$ are doubly degenerate. Then
the thermodynamic potential in presence of strong magnetic field $H(>H^{(c)}$
, critical value discussed later) is given by
\begin{equation}
\Omega_i=-\frac{g_iq_iHT}{4\pi^2}\int d\varepsilon_i\sum_{\nu}\frac{dp_i}
{d\varepsilon_i}\ln [1+exp(\mu_i-\varepsilon_i)/T].
\end{equation}
Integrating by parts and substituting
\begin{equation}
p_{i}=\pm \sqrt{\varepsilon_i^2-m_i^2-2\nu q_i H},\label{eq:en}
\end{equation}
for all T, one finds
\begin{equation}
\Omega_i=-\frac{g_iq_iH}{4\pi^2}\int d\varepsilon_i\sum_{\nu}\frac{2\sqrt{
\varepsilon_i^2-m_i^2-2\nu q_iH}}{[exp(\mu_i-\varepsilon_i)/T+1]}
\end{equation}
where the sum over $\nu$ is restricted by the condition $\varepsilon>
\sqrt{m^2+2\nu q H}$ and the factor 2 takes into account the freedom
of taking either sign in eq(\ref{eq:en}).
For T =0, therefore, approximate the Fermi distribution by a step function
and interchange the order of the summation over $\nu$ and integration
over $\varepsilon$,
\begin{eqnarray}
\Omega_i&=&-\frac{2g_iq_iH}{4\pi^2}\sum_{\nu}\int_{\sqrt{m_i^2+2\nu q_i
H}}^{\mu}
d\varepsilon_i\sqrt{ \varepsilon_i^2-m_i^2-2\nu q_iH} \nonumber\\
&=&-\frac{2g_iq_iH}{8\pi^2}\sum_{\nu}(\mu_i\sqrt{\mu_i^2-m_i^2-2\nu q_i H}
\nonumber \\
& &-(m_i^2+2\nu q_i H)\ln[\frac{\mu_i+\sqrt{\mu_i^2-m_i^2-2\nu q_i H}}
{\sqrt{{m_i}^2 +2\nu q_i H}}]) \label {eq:om}
\end{eqnarray}
Since the temperature $T<<\mu$ at the core of neutron star, the presence
of anti-particles can be ignored. Now instead of infinity the upper limit
of $\nu$ sum can be obtained from the following relation
\begin{equation}
{p_{Fi}}^{2}={\mu_i}^2-{m_i}^2-2\nu q_iH \ge 0,
\end{equation}
where $p_{Fi}$ is the Fermi momentum of the species $i$, which gives
\begin{equation}
\nu \le \frac{{\mu_i}^2-{m_i}^2}{2q_iH}={\nu_{max}}^{(i)}~~(nearest~~integer).
\end{equation}
Therefore, the upper limit is not necessarily same for all the components.
As is well known, the energy of a charged particle changes
significantly in the quantum limit if the magnetic field strength
is equal to or greater than some critical value $H^{(c)} =
{m_i}^2c^3/(q_i\hbar)$ (in G), where $m_i$ and $q_i$ are
respectively the mass and the absolute value of charge of
particle $i$, $\hbar$ and $c$ are the reduced Planck constant
and velocity of light respectively, both of which along with
Boltzman constant $k_B$ are taken to be unity in our choice of
units. For an electron of mass 0.5 MeV, this critical field as
mentioned above is $\sim 4.4\times 10^{13}$G, whereas for a
light quark of current mass 5 MeV, this particular value becomes
$\sim 4.4\times 10^{15}$G, on the other hand for s-quark of
current mass 150 MeV, it is $\sim 10^{19}$G, which is too high
to realise at the core of neutron star. Therefore, the quantum
mechanical effect of neutron star magnetic field on s-quark has been
ignored\cite{cha}. But in the present work, we have considered the
possibility of having with and without the effect on s-quark by
the presence of magnetic field. If the motion of s-quarks are
not effected by the presence of strong magnetic field, the
thermodynamic potential for this component for $T=0$ is given by\cite{far,alc}
\begin{equation}
\Omega_s=-\frac{1}{4\pi^2}[\mu_s\sqrt{{\mu_s}^2-{m_s}^2}
({\mu_s}^2-2.5{m_s})+1.5
{m_s}^4\ln(\frac{\mu_s+\sqrt{{\mu_s}^2-{m_s}^2}}{m_s})].
\end{equation}
In our study, we assume that strange quark matter is charge
neutral and also chemical equilibrium, then
\begin{equation}
\mu_d=\mu_s=\mu=\mu_u+\mu_e, \label{eq:ch}
\end{equation}
and charge neutrality conditions gives
\begin{equation}
2n_u-n_d-n_s-3n_e=0. \label{eq:cha}
\end{equation}
The baryon number density of the system is given by
\begin{equation}
n_B=\frac{1}{3}(n_u+n_d+n_s) \label{eq:ba}.
\end{equation}
Using above eqs(\ref{eq:ch}, \ref{eq:cha}, \ref{eq:ba}) one can solve
numerically for the chemical potentials of all the flavours and
electron.
Now, for $T=0$, we have the number density of the species $i$
($u,~d,~s,~e$)
\begin{equation}
n_i=\frac{g_iq_iH}{4\pi^2}\sum_{\nu}\sqrt{{\mu_i}^2-{m_i}^2-2\nu q_iH}.
\end{equation}
The number density of s-quark in absence of magnetic field is given by
\begin{equation}
n_s=\frac{({\mu_s}^2-{m_s}^2)^{3/2}}{\pi^2}.
\end{equation}
In fig. (1), we have plotted strangeness fraction ($n_s/n_B$) as
function of baryon density. Curve (a) is for low magnetic
field($H=10^{14}$ G) and curve (b) is for with high magnetic field(
$H=10^{18}$ G). Whereas, the
curve (c) is for with high magnetic field($H=10^{18}$ G) without
the presence of magnetic field in s-quark.
In curve (b) and (c), the strangeness fraction is showing an oscillating
behaviour as consecutive  Landau levels are passing the Fermi level.
However, in curve (a), there is no oscillating behaviour because of low
magnetic field.
\par
The total energy density and the external pressure of the
strange quark matter is given respectively by
\begin{eqnarray}
\varepsilon &=& \sum_{i}\Omega_i +B +\sum_{i}n_i \mu_i \nonumber \\
p&=&-\sum_i\Omega_i-B,
\end{eqnarray}
where $i=~u,~d,~s,~e$. Here, we have considered the conventional
bag model for the sake of simplicity in presence of magnetic field.
We are assuming that
quarks are moving freely (non-interacting) within the system and as usual the
current masses of both u and d quarks are extremely small, e.g.,
5 MeV each, whereas, for s-quark the current quark mass is to be
taken 150 MeV. We choose the bag pressure $B$ to be 56
$MeV~fm^{-3}$. Also, we set the magnetic field to be
$H=10^{14}$ G for low and $H=10^{18}$ G for high fields in
our calculations.
\par
We have shown the variation of pressure with energy density in
fig. (2), which is equation of state of strange quark matter.
The curve (a) is for low magnetic field and curve (b) is for
high magnetic field. Whereas, curve (c) is for high magnetic
field without the presence of magnetic field in s-quark.
The curves (b) and (c) show little oscillating behaviour because
of high magnetic field and curve (a) is smooth due to low magnetic
field.
\par
 From the studies of quark matter\cite{alc}, it is predicted that the
mass ($M$) of quark star $\simeq M\odot$ ($M\odot$ solar mass) and
radius ($R$) $\simeq 10$ km. These so-called quark stars have
rather different mass - radius relationship than neutron star
but for stars of mass $\simeq 1.4M\odot$, the structure
parameters of quark stars are very similar to those of neutron stars.
\par
The mass and radius for nonrotating strange quark stars are
obtained by integrating the structure equations of a
relativistic spherical static star composed of a perfect fluid
which is derived from Einstein equation. These equations are
given in ref.\cite{pks}. For a given equation of state, and given
central density, the structure equations are integrated
numerically with the boundary conditions $m(r=0)=0$, to give $R$
and $M$. Though, the equation of states have little oscillating behaviour,
this fact does not effect to characteristic structures of strange stars.
The radius $R$ is defined by the point where $p\simeq
0$. The total gravitational mass $M$, moment of inertia $I$,
surface red shift $z$ and the period $P_0$ corresponding to fundamental
frequency $\Omega_0$ are then given by
$M=m(R),~ I=I(R),~ z=(1-2GM/Rc^2)^{-1/2}-1$ and $P_0=\frac{2\pi}{\Omega_0}$
respectively,
where $\Omega_0=\frac{3GM}{4R^3}$\cite{cut}.  These
are presented in Table 1. Figure 3 shows the variation of mass
with central density for three equation of states as illustrated
in Fig. 2. We noticed from this figure that with increase in magnetic
field strength, the star becomes more compact. The mass and radius
decrease from 2.26 to 1.86 solar mass and from 11 km to 10 km. Similarly,
the values for surface red shift, moment of inertia and
fundamental period decrease, but the central density
increases with magnetic field as we tabulated in table.
\par
In conclusion, we conclude that the presence of strong magnetic
field in strange quark matter reduces the mass and radius of
strange star. That is the star becomes more compact.
Also, the strangeness fraction increases on the average, though, there is
little oscillation due to Landau levels are passing the Fermi level.

\vfil
\eject
\newpage
\begin{table}
\caption {The radius ($R$), mass ($M$), surface red shift ($z$), moment
of inertia ($I$) and period of fundamental frequency ($P_0$) of strange
stars versus central density $\rho_c$
for three cases; (a): $H=10^{14} $ G, (b): $H=10^{18}$ G and
(c): $H=10^{18}$ G but absence of magnetic field in s-quark.}
\hskip 0.5 in
\begin{tabular}{ccccccccc}
\hline
\multicolumn{1}{c}{$\rho_c$} &
\multicolumn{1}{c}{$R$}&
\multicolumn{1}{c}{$M/M_{\odot}$}&
\multicolumn{1}{c}{$z$}&
\multicolumn{1}{c}{$I$}&
\multicolumn{1}{c}{$P_0$}&
\multicolumn{1}{c}{$Cases$}\\
\multicolumn{1}{c}{($g~cm^{-3}$)} &
\multicolumn{1}{c}{($km$)}&
\multicolumn{1}{c}{} &
\multicolumn{1}{c}{} &
\multicolumn{1}{c}{($g~cm^{-2}$)} &
\multicolumn{1}{c}{($ms$)} &
\multicolumn{1}{c}{}\\
\hline
2.5$\times 10^{15}$&11.22&2.26&0.57&3.50$\times 10^{45}$&0.49&(a)\\
3.5$\times 10^{15}$&10.01&1.86&0.49&2.41$\times 10^{45}$&0.46&(b)\\
3.5$\times 10^{15}$&9.85&1.83&0.48&2.13$\times 10^{45}$&0.45&(c)\\
\hline
\end{tabular}
\end{table}
\vfill
\eject
\newpage
\begin{figure}
\caption { Strangeness fraction and baryon density curves for three cases;
Case I: $H=10^{14}$ G (curve a); Case II: $H=10^{18}$ G (curve b) and
Case III: $H=10^{18}$ G (curve c) but absence of magnetic field in s-quark.}
\end{figure}
\begin{figure}
\caption { Pressure and energy density curves for three cases as mentioned
in figure 1.}
\end{figure}
\begin{figure}
\caption { Mass and central density curves for three cases as mentioned
in figure 1.}
\end{figure}
\vfil
\eject
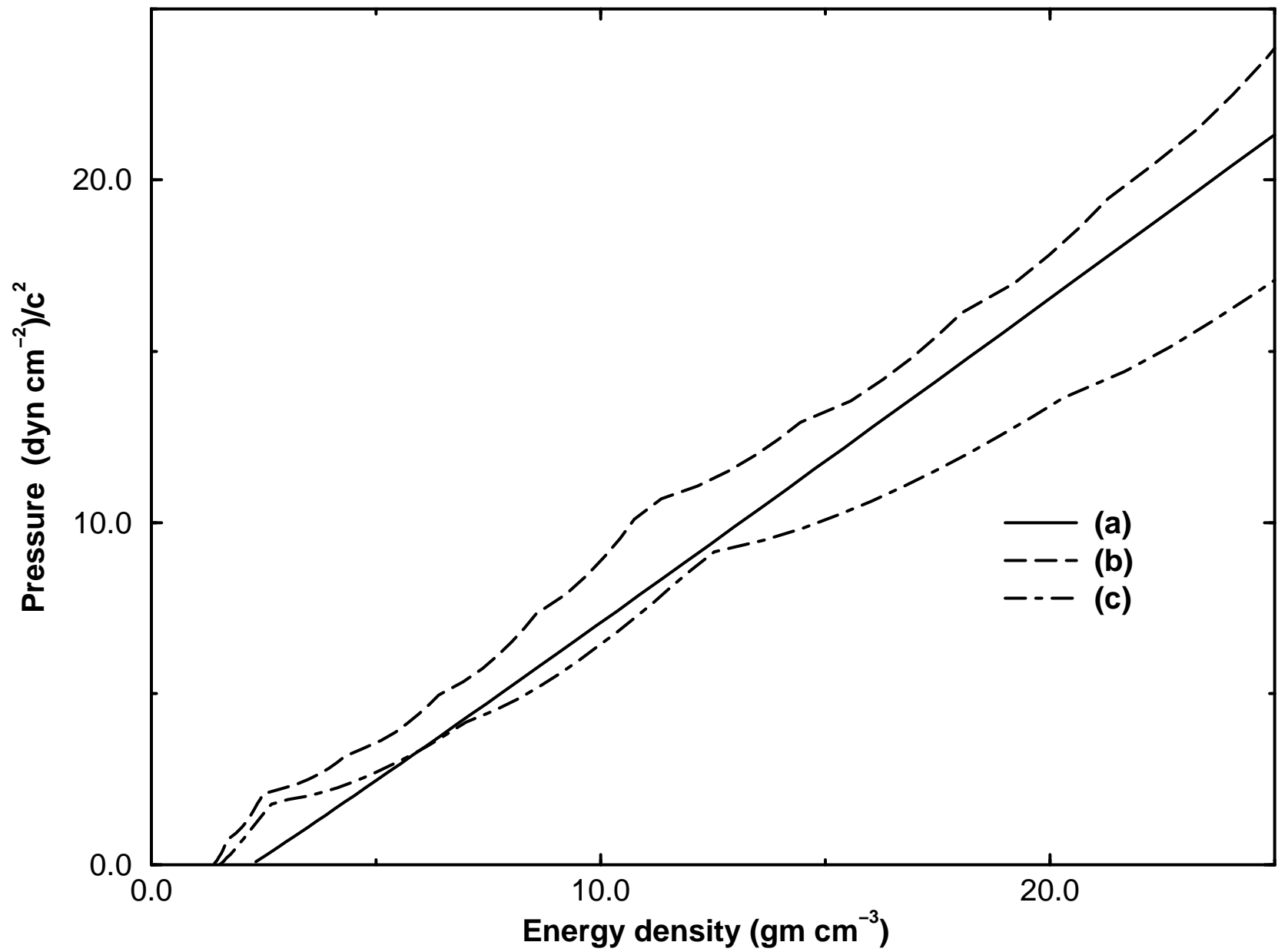
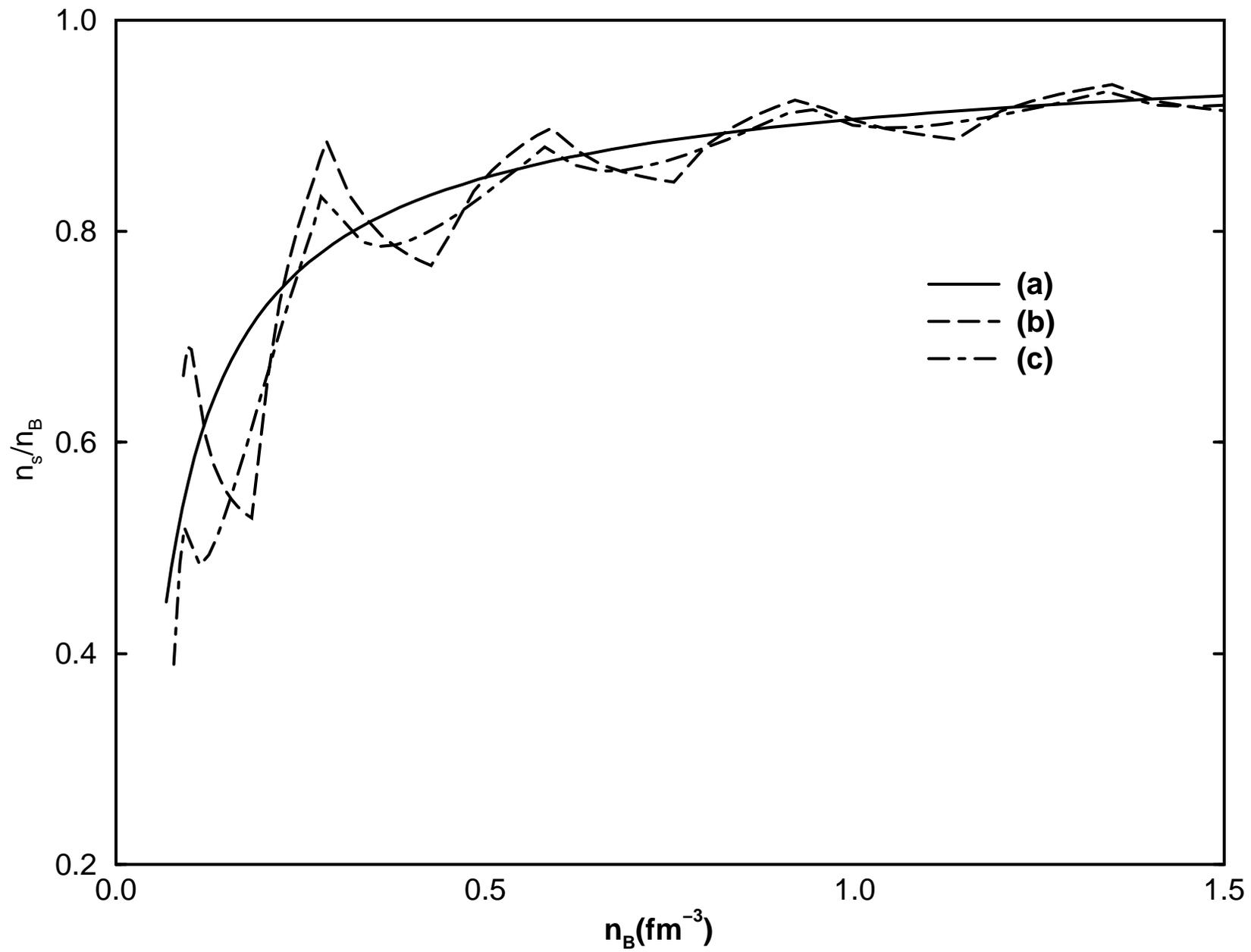
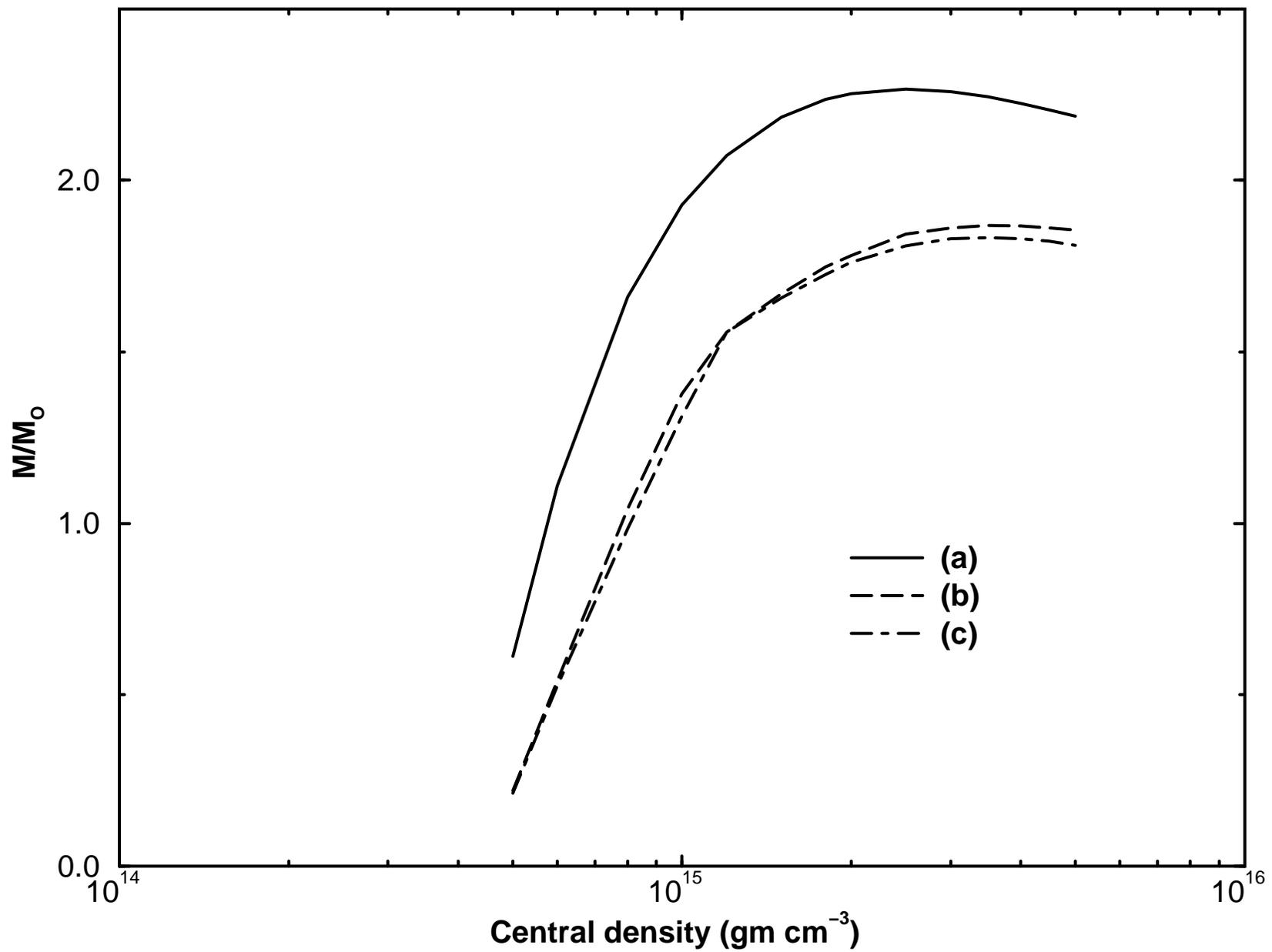
\end{document}